\documentclass[twocolumn,prl,showpacs]{revtex4}  

\usepackage{graphicx}

\usepackage{epsfig}

\begin{document}

\title {New kind of dodecagonal quasicrystal}

\author{Alfredo Metere$^{1c}$, Peter Oleynikov$^1$, Mikhail Dzugutov$^2$ and Sven Lidin$^3$}

\affiliation{ $^1$ Department of Materials and Environmental Chemistry,
  Stockholm University, Arrhenius V\"{a}g. 16C S-10691 Stockholm,
  Sweden\\ $^2$ Department of Mathematics, Royal
  Institute of Technology, SE-100 44 Stockholm, Sweden\\ $^3$Division of Polymer \& Materials Chemistry Lund University
  SE-221 00 Lund Sweden}

\begin{abstract}
  We report a novel kind of dodecagonal quasicrystal that has so far
  never been observed, nor theoretically predicted. It is composed of
  axially stacked hexagonal particle layers, with 12-fold rotational
  symmetry induced by $30^{\circ}$ rotation of adjacent layers with
  respect to each other. The quasicrystal was produced in a
  molecular-dynamics simulation of a single-component system of
  particles interacting via a spherically-symmetric potential, as a
  result of a first-order phase transition from a liquid phase under
  constant-density cooling.  This finding implies that a similarly
  structured quasicrystal can possibly be produced by the mesogens of
  the kind that produce smectic-B crystals and in a system of
  spherically-shaped colloidal particles with appropriately tuned potential.
  \end{abstract}

\date{\today}
% 61.44.Br Quasicrystals, Structure of
% 61.44.-n Structure, of quasicrystals
% 83.10.Rs Molecular Dynamics, computer simulation of


\maketitle

Quasiperiodic order in mesoscopic soft-matter systems is now a new
research frontier of rapidly growing interest \cite{DUB}.  First of
all, the interest in the studies of mesoscopic-scale soft-matter
quasicrystals is motivated by the perspective application of
mesoscopic quasicrystals in photonics \cite{VARD}.  Following the
first discovery of a dodecagonal quasicrystal in a micelle-forming
system \cite{UNG}, similar structures were produced in colloids
\cite{FISCH}, mesoporous silica \cite{CHAN}, binary system of
nanoparticles \cite{TAL} and star polymers \cite{DO2}. Soft-matter
quasicrystals were also simulated using molecular dynamics
\cite{GLOTZ}. However, the structure of all these quasicrystals
reproduces, on mesoscopic scale, the tetrahedrally close-packed
structure of the quasicrystals observed in intermetallic
compounds. One independent way to induce quasiperiodic order to be
mentioned is superposition of Faraday waves \cite{RON}.

In this context, mesogenic systems forming smectic phases
\cite{CHANDR} are of particular interest. These phases, usually
composed of elongated particles, exhibit layered structures with
uniaxial particle orientation \cite{CHANDR, FRE}. So far, the only
kind of quasiperiodic order observed in smectic phases has been twist
grain boundaries (TGB) structure, where the layers in adjacent
structural blocks are commensurately rotated by an appropriate angle
around a helical axis parallel to the layers \cite{LUB, GOOD}.

Smectic mesophases commonly solidify into the 6-fold smectic-$B$
crystal which represents a uniaxial stacking of coherently
oriented hexagonally close-packed layers \cite{LE1,FRE}. Here, we
report a novel type of solid smectic phase possessing 12-fold
symmetry produced in a molecular-dynamics simulation of a simple
one-component system. Like the smectic-B crystal, it consists of
hexagonally ordered layers axially stacked in $ABA$ order, but in
contrast to the crystal, its $A$ and $B$ layers are rotated by
$30^{\circ}$ with respect to each other, in the layer plain. This
is a new  smectic solid phase, and a new type of
quasicrystal that has never been observed in other systems, nor
theoretically predicted.

The molecular-dynamics model we utilized in this study consists
of 16384 identical particles confined to a cubic box with
periodic boundary conditions. The interparticle interaction was
assumed to be spherically-symmetric, described by the pair
potential  (see Methods).

We investigated the system's phase behaviour under temperature
variation at the constant number density $\rho=0.32$.  The
temperature was changed in a stepwise manner, with comprehensive
equilibration of the system at a new temperature after each step.
At the beginning, the system was equilibrated in its
thermodynamically stable isotropic liquid state at sufficiently
high temperature, after which it was subjected to isochoric
cooling.

Fig. 1 presents the system's energy and pressure variations as
functions of temperature at the number density $\rho=0.32$. A
discontinuous change in the thermodynamic parameters was detected
as the liquid had been cooled below $T=1.1$. It was also found
that this thermodynamic singularity was accompanied by a sharp
drop in the rate of self-diffusion. These discontinuities
convincingly demonstrate that the system performed under cooling
a first-order phase transition from the liquid state to a solid
phase. This conclusion can be confirmed by a significant
hysteresis that was observed when re-heating the low-temperature
phase, see Fig. 1. A non-trivial character of the produced
low-temperature solid phase was indicated by an anomalously long
time required for its equilibration which amounted to about a
billion of time-steps.

\begin{figure}[h] %figure 1

\includegraphics[width=8cm]{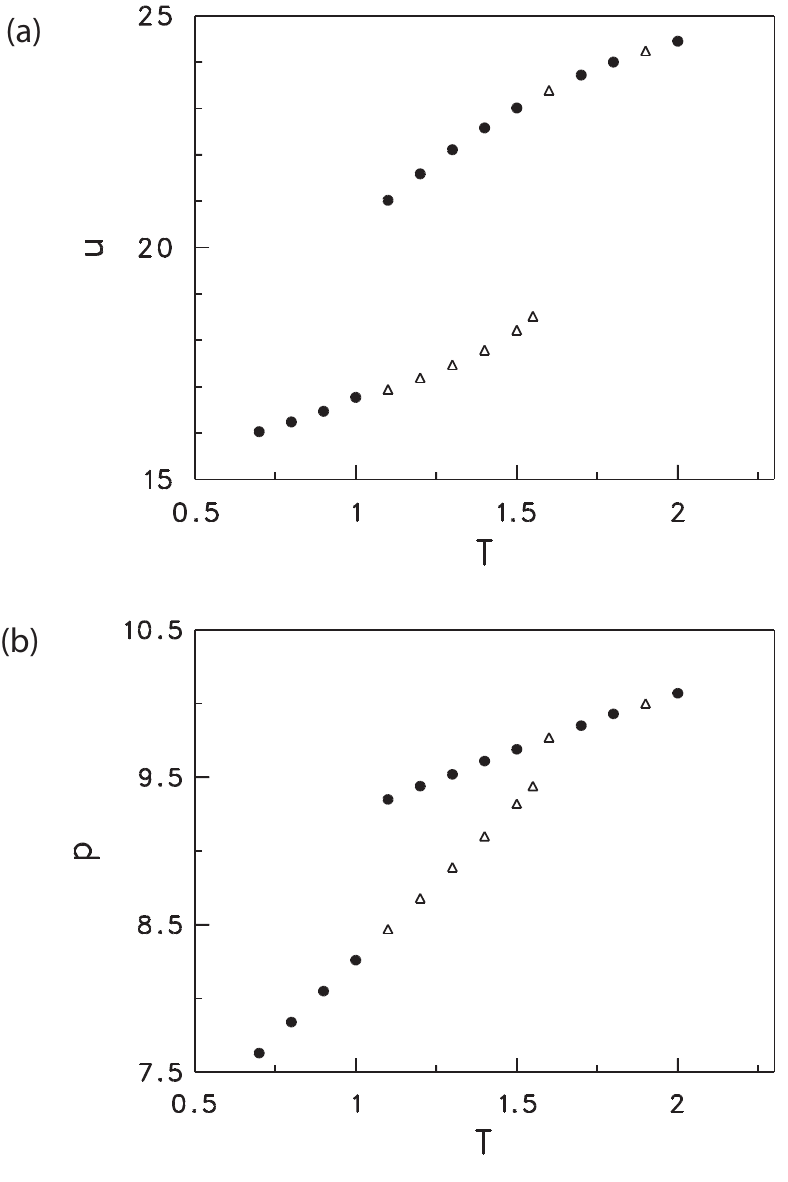}
\caption{ Liquid-solid phase transformation. $(a)$ and $(b)$,
  respectively: the temperature variation of the energy and pressure
  at the number density $\rho = 0.32$. Dots: cooling; triangles:
  heating. }
\label{fig1}
\end{figure}

\begin{figure} [h]%figure 2
\includegraphics[width=8cm]{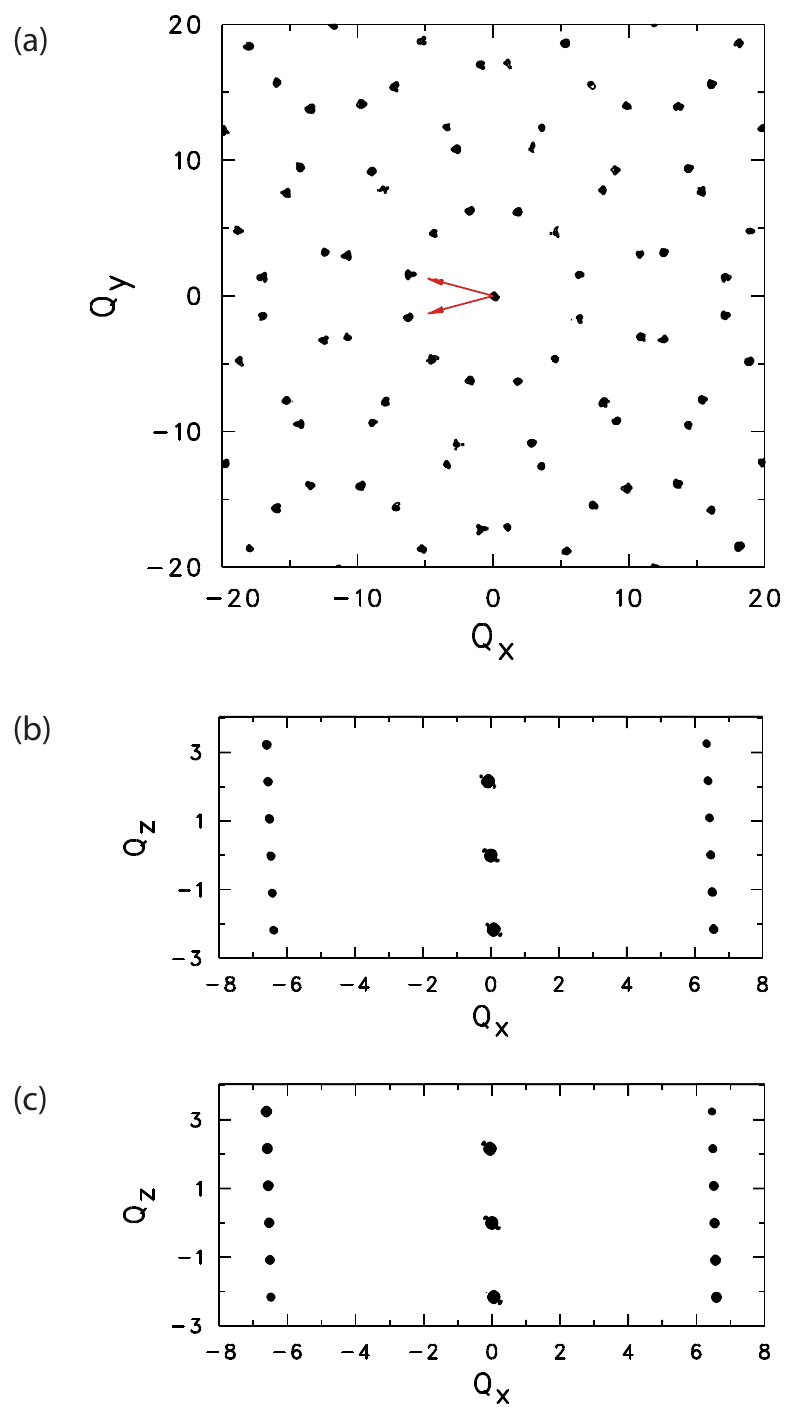}
\caption{ Isointensity contour plots of the structure factor
  $S({\bf Q})$ (see Methods). $(a)$: in the {\bf Q}-plane
  perpendicular to the axis. $(b)$ and $(c)$ represent respectively two {\bf Q}-planes defined by the axis and the two vectors, $30^{\circ}$ apart, shown in the top panel.}
\label{fig2}
\end{figure}

\begin{figure} [h] %figure 3
%\centering
\includegraphics[width=8cm]{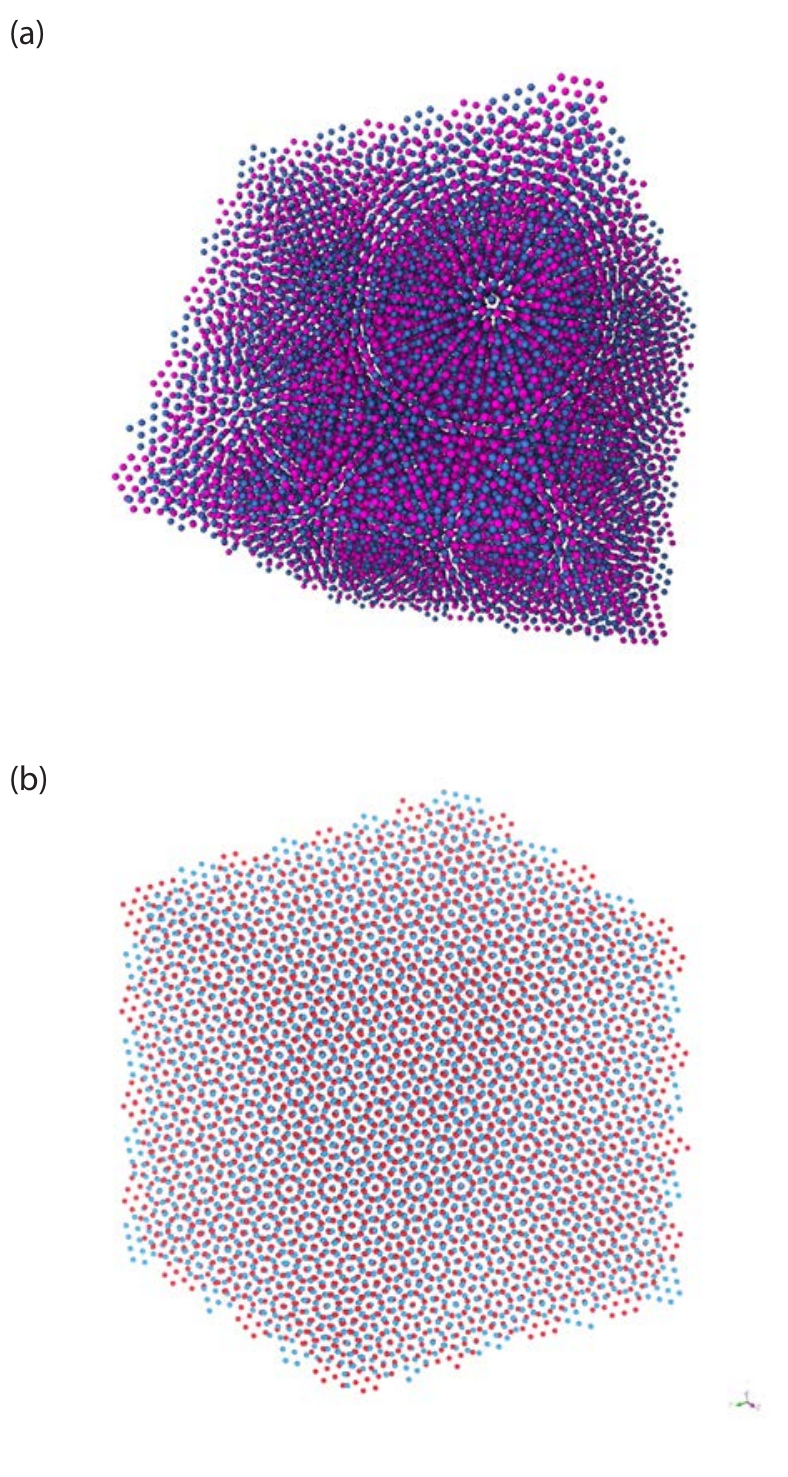}
\caption{ An axial view of the configuration. $(a)$: perspective
  projection. $(b)$: orthogonal projection. Adjacent A and B
  layers are discriminated by color.\\.}
\label{fig3}
\end{figure}
\begin{figure} %figure 4
\includegraphics[width=8cm]{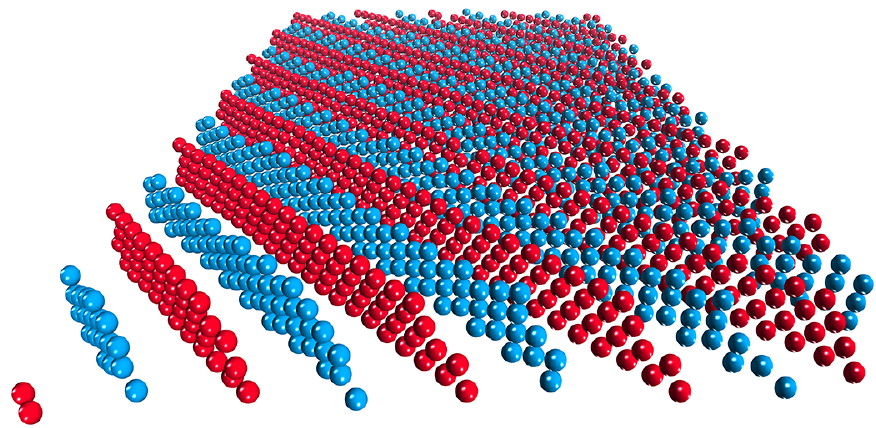}
\caption{  A slice of the configuration cut parallel to the axis.}
\label{fig4}
\end{figure}

\begin{figure} %figure 5
\includegraphics[width=8cm]{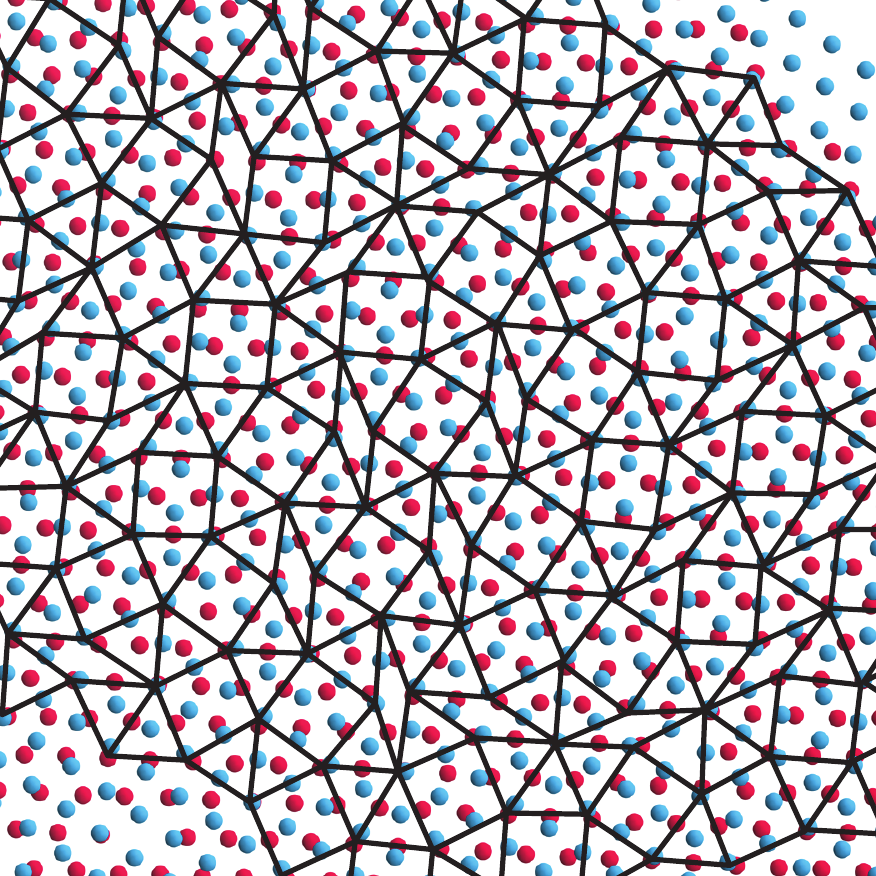}
\caption{Tiling of the axial projection of two adjacent particle
layers, produced by connecting centres of adjacent
12-particle rings.}
\label{fig4}
\end{figure}

The structure characterisation of the low-temperature solid phase
has been performed by inspecting the reciprocal-space pattern of
its density distribution. For that purpose, we calculated the
structure factor $S({\bf Q})$ (see Methods) which represents the
scattered beam intensity as measured in the diffraction
experiments. To remove thermally induced fluctuations, the
simulated configuration was subjected to the steepest descent
energy minimisation which mapped it onto the nearest minimum of
its energy landscape. This minimum was found to correspond to the
number density $\rho=0.31$.

As a first step in this structure analysis, we determine the global
symmetry of the configuration, and its axis orientation (see
Methods). The latter having been found, we calculated $S({\bf Q})$ on
the reciprocal-space plane $Q_z=0$, $Q_z$ being the axis coordinate;
this result is presented in Fig. 2a. The pattern of $S({\bf Q})$
maxima observed in that ${\bf Q}$ plane is distinctively 12-fold. To
further demonstrate the 12-fold rotational invariance of the simulated
configuration, we also calculated the structure factor in two planes
defined by the axis and two translational symmetry vectors,
$30^{\circ}$ apart, which are indicated in Fig. 2a. The results, shown
in Fig. 2b, are indeed consistent with the conclusion that the
reciprocal-space image of the simulated structure is invariant with
respect to $30^{\circ}$ rotation.

Besides, the two diffraction patterns in Fig. 2b look like those
observed in typical smectic-B crystals \cite{DZ1,LE1}. They
represent a structure composed of flat close-packed particle
layers axially stacked in $ABA$ order \cite{LE2}. The diffraction
results in Fig. 2b also make it possible to estimate the ratio of
the interlayer spacing and the interparticle distance within a
layer as about $3$, which can be compared with the aspect ratio
of constituent particles in commonly occurring smectic phases. We
note that this value is is consistent with the ratio of the
positions of two minima in the pair potential, see Fig. 1a. Thus,
the configuration is structurally similar to the smectic-$B$
crystal, except that, in contrast to hexagonal symmetry of the
latter, it exhibits 12-fold symmetry.

These diffraction results can be interpreted by assuming that the
system's apparent 12-fold axial symmetry is induced by
$30^{\circ}$ rotation of the $A$ and $B$ subsystems of layers
with respect to each other, in the layer plane. This conclusion
is supported by the visual inspection of the real-space images of
the simulated structure presented in Fig. 3. The view from the
axial direction, Fig. 3a, shows a pattern of 12-particle rings
characteristic of dodecagonal structures \cite{GA, DZ2}, and its
particle density lines that can be detected by looking at the
grazing angle exhibit $30^{\circ}$ rotational invariance. When
viewed along the layer plane, Fig. 3b, the configuration looks
like a typical smectic-B crystal with $ABA...$ layer stacking;
its only distinction from the latter in that the adjacent $A$ and
$B$ layers are rotated with respect to each other, in the layer
plane, by $30^{\circ}$. This can also be seen in Fig. 3c.

In Fig. 4 we present tiling of the axial projection of two
adjacent layers of the quasicrystal. The tiling is produced by
connecting the centres of adjacent apparent 12-particle rings
which the pattern is composed of. Besides equilateral triangles
and squares, typical for periodic tilings, the present tiling
includes the $30^{\circ}$ rhombus, an element specific for a
12-fold tiling pattern \cite{GA, DZ2}.

In crystallographic classification, the symmetry of this quasicrystal
is $P12_6/mmc$. The quasicrystal is composed of stacked layers of
$6^3$ nets that each have planar $6mm$ symmetry.  The stacking along
the axis with a $\pi/6$ relative rotation of the adjacent layers
preserve the mirror operations and subsequent layers are related by a
$12_6$ screw operation and a diagonal glide reflection with a $1/2c$
component. We note that the screw operation producing this structure
results in extinction of some reflections in its diffraction pattern
which are characteristic of the diffraction patterns of conventional
 tetrahedrally close-packed dodecagonal quasicrystals, e.g. those found in
intermetallic systems \cite{ISHI}. In this respect, it should be mentioned that
the diffraction pattern of the present structure, Fig. 2a, is
identical to that of the twinned multidomain fcc structure produced by
staking of $6$-fold domains alternate $0^{\circ}, 30^{\circ}$-rotated
in $111$ plane \cite{FISCH}. In this way, the present structure is
conceptually similar to the quasicrystalline TGB phases which too give
rise to the diffraction patterns that differ significantly from those
observed in conventional quasicrystals.

Thus, we found a new type of smectic quasicrystaline phase,
alternative to the TGB quasicrystals. The new phase differs from
the latter in real-space structure while resembling it in terms
of general concept. The new quasicrystal can possibly be
classified as a {\it dodecagonal smectic-B} phase.

A conceptually significant peculiarity of the simulated smectic
quasicrystal is that it exhibits complete absence of local phason
disorder. That the layers are perfectly identical can be
concluded by inspecting the general view of the configuration
from the axial direction which is presented in Figure 3. This
feature appears to be a generic property of this quasicrystal due
to its physical nature, and the way of its layer stacking. Due to
the strong interaction between particles within a layer induced
by the first minimum in the potential, Figure 1, the hexagonal
layer structure is solid, making the energy cost for intralayer
dynamics forbiddingly high. Thus, this quasicrystal structure
represents a single energy minimum, which excludes any
possibility for energetically degenerate local phason
flips\cite{DZ3}. The absence of phason dynamics, and its
respective entropic contribution to the free energy
\cite{HEN}implies that the mechanism of thermodynamic stability
of this quasicrystal is entirely energetic, which suggests that
this structure might be considered as a candidate for the
quasiperiodic ground state.

On the other hand, this configuration, constrained by periodic
boundary conditions, is a periodic approximant, and the absence
of local phason dynamics implies the presence of a uniform phason
strain which can be observed both in the diffraction pattern and
by inspecting the density lines in the real-space configuration,
Fig. 3a, at a grazing angle. It arises from the fact that each of
the two subsets of 6-fold periodic layers must independently be
oriented consistently with the periodic boundaries. Therefore,
the rotation angle between the two cannot be exactly
$30^{\circ}$, and the deviation is size-dependent.

Two possible ways of materialization of this structure in real
physical systems can be conjectured. First, the morphological
similarity of the simulated smectic quasicrystal and the
smectic-B crystal strongly suggests that some mesogens of those
forming the latter phase may also be able to freeze into a
quasiperiodic structure like the one we report here. Moreover,
these mesogens may also form a liquid-crystal ``dodecatic''
mesophase that was recently suggested \cite{DO} as a dodecagonal
counterpart of the 6-fold hexatic mesophase. It is suggested to
possess 12-fold symmetry both in the local positional order and
in its global bond-orientational order.

Second, the main features of the spherically-symmetric
interparticle potential used in the present model show similarity
to the force field predicted for colloidal systems by the
classical Deryagin-Landau-Verwey-Overbeek theory \cite{VO,IS}
(amended with hard core repulsion or steric repulsion at close to
contact). This suggests a possibility that smectic-like layered
structures exhibiting dodecagonal symmetry similar to the one
reported here can be produced in colloidal systems of spherically
shaped particles, with appropriate tuning of the effective potential.

In summary, we report a molecular-dynamics simulation
demonstrating that a system of identical particles interacting
via a spherically-symmetric potential can form a phase
morphologically similar to the smectic-$B$ crystal but possessing
12-fold symmetry. This is a new kind of smectic quasicrystal,
alternative to the quasicrystalline TGB phases which it resembles
in terms of the general design principle. This finding opens a
perspective of producing similar quasicrystals in the mesogens
that are known to form the smectic-B crystals. Conceptually, it
introduces a new formation mechanism of quasiperiodic order. It
also changes the basic model of smectic phases, thereby advancing
our understanding of the causes underlying the occurrence of
particular structures in the phase
transformations of liquid crystals.\\

\section {Appendix: Methods}

\subsection {Pair potential}

The pair potential used in this simulation is shown in Fig. 5.

\begin{figure}[h] %figure 6
\includegraphics[width=8cm]{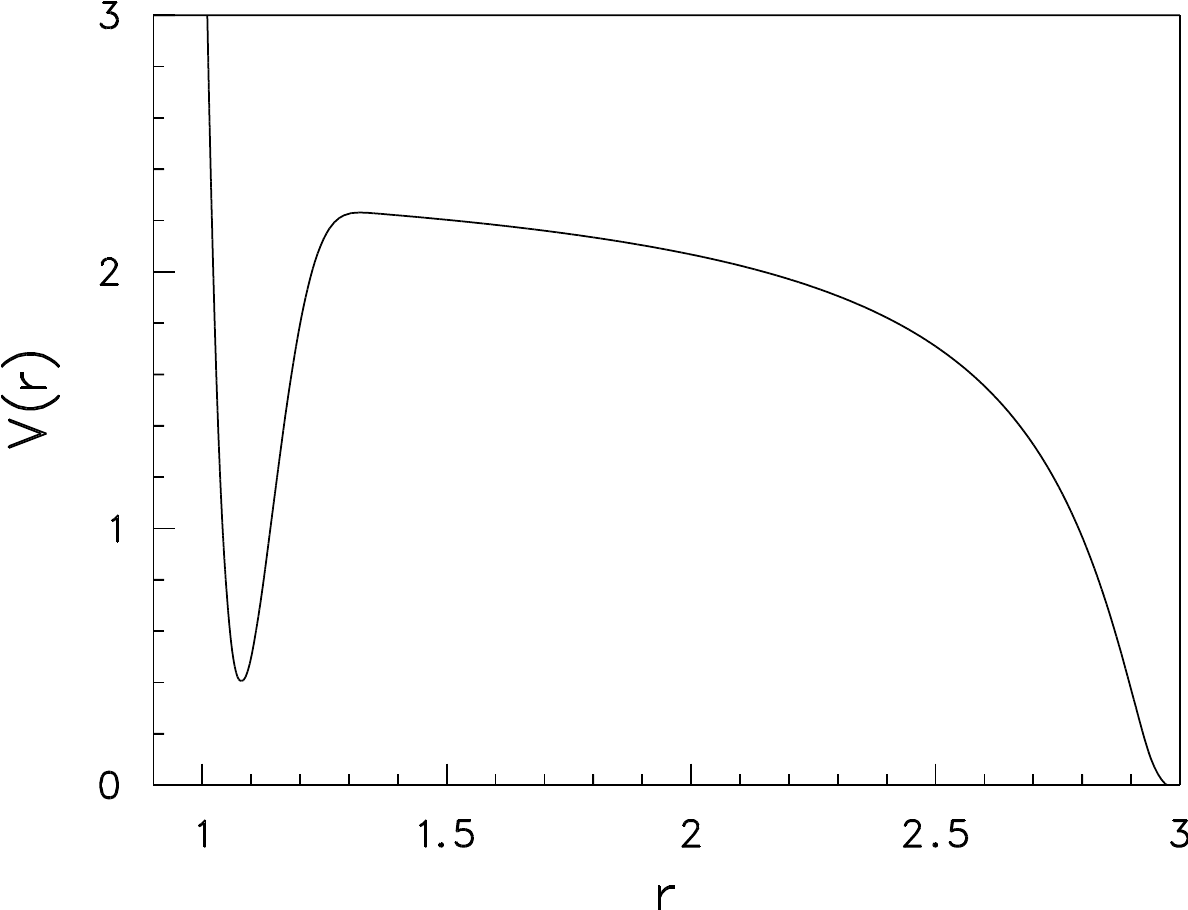}
\caption{Pair potential}
\label{fig1}
\end{figure}

The functional form of the potential energy for two particles
separated by the distance $r$ is:

\begin{equation} 
V(r)  = a_1  ( r^{-m} - d) H(r,b_1,c_1) + a_2 H(r,b_2,c_2)  \\ 
\end{equation}
where
\begin{equation} 
H(r,b,c) = \left\{ \begin{array}{ll}
  \exp\left( \frac{b}{r-c} \right) & r < c\\
0 & r \geq c
\end{array}
\right.       
\end{equation}

\begin{table} [h]
\begin{tabular}{cccccccc}

\hline 
\hline 

m & $a_1$ & $b_1$ & $c_1$ & $a_2$ & $b_2$ & $c_2$ & $d$ \\

\hline 
12 \space & 265.85 \space & 1.5 \space & 1.45 \space & 2.5 \space & 0.19 \space &
3.0 \space & 0.8\\

\hline \\
\end{tabular}
\caption{Values of the  parameters for the pair potential.} 
\label{table1}
\end{table}

The values of the parameters are presented in Table I. The first
term describes the short-range repulsion, and the minimum,
whereas the second term expresses the long-range repulsion. All
the quantities we report here are expressed in terms of the
reduced units that were used in the definition of the
potential. Note that its short-range repulsion part, and the
position of the minimum, are consistent with those of the
Lennard-Jones (LJ) potential \cite{HAN}, which makes it possible
to compare the reduced number densities of the two systems.

This pair potential represents a modification of an earlier reported
pair potential \cite{DZ1} that was found to produce the smectic-B
crystal. The main difference between these two potentials is that in
present one the long-range repulsion branch is extended to a
significantly larger distance. This modification was intended to
increase spacing between the layers, thereby reducing the interlayer
cohesion.

\subsection {Structure characterisation}
\noindent
In the reciprocal space, the structure of a simulated system of
particles is presented in terms of static structure factor $
S({\bf Q})$ that is defined as
\begin{equation}
  S({\bf Q})= \langle \rho({\bf Q})\rho(-{\bf Q})\rangle 
\end{equation}
where $\rho({\bf Q})$ is the ${\bf Q}$-component of the spatial
Fourier-transform of the instantaneous number density distribution of
a system of $N$ particles:
\begin{equation}
\rho({\bf Q},t) = \frac{1}{\sqrt{N}} \sum_{i=1}^{N} \exp [-i{\bf Q r}_i]
\end{equation}
${\bf r_i}$ being the position of particle $i$, and $\langle \
\rangle$ denoting  ensemle averaging.\\
A spherically-averaged static structure factor can be calculated
\cite{HAN} from the spherically invariant radial distribution
function $g(r)$ as:
\begin{equation}
  S(Q) = 1 + 4 \pi \rho \int_{0}^{\infty} [g(r)-1] \frac{\sin(Qr)}{Qr} r^2 dr
\end{equation}
As a first step in this structure analysis, we calculated $S({\bf Q})$
on the reciprocal-space sphere of the radius corresponding to the
position of the first peak of the spherically averaged structure
factor (see Methods). That calculation produced a pattern of
well-defined $S({\bf Q})$-maxima, which represent decomposition of the
peak of the spherically averaged structure factor  on thus defined
$\bf Q$-sphere. This made it possible to determine the global symmetry
of the configuration, and the axis orientation.

\end{document}